\documentclass[twocolumn,9pt]{article} % here we use the article class, rather than elsarticle

\usepackage[square,numbers,sort&compress,comma]{natbib}

\usepackage{amsmath}
\usepackage{amssymb}
\usepackage{caption}
\usepackage{tabularx}
\usepackage{graphicx}
\usepackage{latexsym}
\usepackage{times}
\usepackage{soul}
\usepackage{xcolor}
\sethlcolor{yellow}
\usepackage[pagewise]{lineno}
\usepackage{hyperref}

\topmargin - 12pt % might need to be set to 0pt for some installations
\renewenvironment{abstract}%
              {% - begin definition
               \small% - select font
               {\bfseries \abstractname}% - select font
               \par% - end a paragraph (skip \parsep)
               \vspace{10pt}% - add vertical space
              }% - complete definition

\renewcommand\abstractname{Abstract}

\newcommand{\nomenclature}% - name of command
              [1]% - number of arguments
              {% - begin definition
               \bgroup% - begin a local group
               \flushleft% - turn on flushleft option
               \small\bf% - select font
               #1% - insert title text
               \par% - end a paragraph (skip \parsep)
               \egroup% - terminate local group
              }% - complete definition

\renewcommand{\section}% - name of command
              [1]% - number of arguments
              {% - begin definition
               \bgroup% - begin a local group
               \flushleft% - turn on flushleft option
               \small\bf% - select font
               \refstepcounter{section}% - increment counter
               \arabic{section}. #1% - insert title text
               \par% - end a paragraph (skip \parsep)
               \egroup% - terminate local group
              }% - complete definition

\renewcommand{\subsection}% - name of command
              [1]% - number of arguments
              {% - begin definition
               \bgroup% - begin a local group
               \flushleft% - turn on flushleft option
               \small\em% - select font
               \refstepcounter{subsection}% - increment counter
               \arabic{section}.% - insert title text
               \arabic{subsection}. #1% - insert title text
               \par% - end a paragraph (skip \parsep)
               \egroup% - terminate local group
              }% - complete definition

\renewcommand{\subsubsection}% - name of command
              [1]% - number of arguments
              {% - begin definition
               \bgroup% - begin a local group
               \flushleft% - turn on flushleft option
               \small\em% - select font
               \refstepcounter{subsubsection}% - increment counter
               \arabic{section}.% - insert title text
               \arabic{subsection}.% - insert title text
               \arabic{subsubsection}. #1% - insert title text
               \par% - end a paragraph (skip \parsep)
               \egroup% - terminate local group
              }% - complete definition

  \newcommand{\acknowledgement}% - name of command
              [1]% - number of arguments
              {% - begin definition
               \bgroup% - begin a local group
               \flushleft% - turn on flushleft option
               \small\bf% - select font
               #1% - insert title text
               \par% - end a paragraph (skip \parsep)
               \egroup% - terminate local group
              }% - complete definition

  \newcommand{\sectionbib}% - name of command
              [1]% - number of arguments
              {% - begin definition
               \bgroup% - begin a local group
               \flushleft% - turn on flushleft option
               \small\bf% - select font
               #1% - insert title text
               \par% - end a paragraph (skip \parsep)
               \egroup% - terminate local group
              }% - complete definition

\begin{document}

% -------------------------------------------------------------------- %
% -------------------------------------------------------------------- %
% -------------------------------------------------------------------- %

% -------------------------------------------------------------------- %

\small
\baselineskip 10pt

% -------------------------------------------------------------------- %
% -------------------------------------------------------------------- %
% -------------------------------------------------------------------- %
\setcounter{page}{1}
% -------------------------------------------------------------------- %
\title{\LARGE \bf Evolution of lean hydrogen-air premixed flames under high-frequency acoustic forcing: flame morphology and displacement speed}

\author{{\large Xinyi Chen$^{a,1*}$, Frederick W. Young$^{a,1*}$, Umair Ahmed$^{a}$, Robert Stewart Cant$^{b}$}\\[10pt]
        {\footnotesize \em $^a$School of Engineering, Newcastle University, Newcastle upon Tyne NE1 7RU, United Kingdom}\\[-5pt]
        {\footnotesize \em $^b$Department of Engineering, University of Cambridge, Cambridge CB2 1PZ, United Kingdom}}

\date{}  %%% Leave as is, do not add date;

% -------------------------------------------------------------------- %
% -------------------------------------------------------------------- %
% -------------------------------------------------------------------- %
\twocolumn[\begin{@twocolumnfalse}
\maketitle
\rule{\textwidth}{0.5pt}
\vspace{-5pt}

\begin{abstract} % 100 to 300 words.
Fully compressible numerical simulations of two-dimensional laminar lean hydrogen-air premixed flames have been performed, with the flame front subjected to acoustic forcing through the specification of a monopole-type sound source at the inflow. Simulations have been performed for acoustic frequencies ranging from 35~kHz to 500~kHz at two equivalence ratios, $\phi = 0.4$ and $\phi = 0.7$. During the flame-acoustic interaction, the flame evolves from an initially weakly stretched state to exponential perturbation growth, wrinkle interaction, and the formation of non-linear cellular structures, with distinct linear and non-linear stages identified from Fourier mode analysis. The instability dynamics depend strongly on both forcing frequency and equivalence ratio. In the case of $\phi=0.4$, the flame behaviour is strongly influenced by thermodiffusive instability, with a characteristic sequence of uniform cells, cell splitting, and cell merging. For $\phi=0.7$, weaker thermodiffusive effects result in a response more strongly governed by hydrodynamic instability and large-scale wrinkle growth. At low forcing frequencies, flame corrugations remain relatively uniform, whereas at high frequencies the flame front becomes increasingly modulated and develops envelope-like structures, which can be interpreted as the interaction between an intrinsic standing cellular mode and the imposed acoustic disturbance. In the linear growth regime, the density-weighted displacement speed, $S_d^*$, shows a linear correlation with total stretch rate, $K$, for all forcing frequencies. While in the non-linear growth regime, two distinct branches appear, corresponding to weakly stretched flame segments and strongly negatively curved segments associated with flame pinch-off.
 
\end{abstract}

\vspace{10pt}

{\bf Novelty and significance statement}

\vspace{10pt}

This paper reports results from numerical simulations for lean hydrogen-air premixed flames subjected to acoustic oscillations originating from an approximate acoustic monopole. The study addresses the dependence of flame instability patterns on both acoustic forcing frequency and equivalence ratio for the first time. A conceptual framework is proposed to interpret the behaviour of flame evolution. The correlations between density-weighted displacement speed and the total stretch rate are reported for the linear and non-linear growth phases of the flame for the first time in the case of acoustically excited lean hydrogen-air flames. These findings advance the fundamental understanding of acoustically triggered intrinsic instabilities and provide a basis for active control of hydrogen flames in practical combustion systems.

\vspace{5pt}
\parbox{1.0\textwidth}{\footnotesize {\em Keywords:} lean hydrogen; flame-acoustic interaction; flame instability; Markstein number; displacement speed.}
\rule{\textwidth}{0.5pt}
$^{1}$First author, 
*Corresponding author.
\vspace{5pt}
\end{@twocolumnfalse}] 

% \linenumbers
\section{Introduction\label{sec:introduction}} \addvspace{5pt}

In practical combustion systems such as gas turbines, it has long been recognized that flames can generate or amplify acoustic oscillations through thermoacoustic coupling, where unsteady heat release interacts with pressure fluctuations in accordance with Rayleigh’s criterion~\cite{rayleigh1878explanation}. Consequently, the interaction between flames and acoustic waves constitutes a fundamental problem in reacting flows, and is studied to mitigate the formation of thermoacoustic instabilities. 

Searby and Rochwerger~\cite{Searby1991} 
%identified two distinct instability modes during flame-acoustic interactions, which are commonly referred to as the primary and secondary (thermo-diffusive) acoustic instabilities, analogous to parametric oscillations in other resonant systems. They 
developed an analytical framework that allows the prediction of the parametric stabilisation and resonance thresholds in terms of the amplitude of the acoustic wave and the wave number of the perturbation. Theoretical thresholds have been compared with experimental measurements for premixed flames propagating in tubes under acoustic forcing~\cite{Searby1991}. While this framework shows good agreement for hydrocarbon--air flames~\cite{Searby1991,Denet1995}, %which do not involve strongly diffusive species,
significant discrepancies emerge for lean hydrogen--air premixed flames~\cite{Veiga-Lopez2020}.
The discrepancy mainly arises because the analytical expressions are derived under the assumptions of near-equidiffusive mixtures and large activation energy. These assumptions limit the applicability of the theory and make it unsuitable for lean hydrogen flames, where strong thermodiffusive instabilities and broad reaction zones are present ~\cite{Clavin1982}. The complex instability behaviour of fuel lean hydrogen combustion remains poorly understood and is of increasing relevance given the growing adoption of hydrogen as a carbon-free energy carrier for future sustainable power generation.

During flame-acoustic interaction, the Markstein number plays a central role in linking flame dynamics with acoustic interactions. %Combustion occurs at the molecular scale, 
%as a result of collisions between reacting molecules,
%where molecular transport of heat and mass generally plays a crucial role \cite{Bray1991}. 
%Consequently, combustion can be strongly influenced by events occurring
%at the fluid dynamics scales. 
When the upstream flow is perturbed by an acoustic field, the local burning rate responds to flame curvature and flow non-uniformity \cite{LieuwenT}, with this sensitivity characterized by the Markstein number~\cite{markstein1964}. The resulting fluctuations in heat release feed energy back into the acoustic field, potentially driving parametric oscillations. Building on this mechanism, Markstein number can be included into a Mathieu-type equation to predict thresholds for parametric resonance and stabilization~\cite{Searby1991}. Therefore, assessing the flame sensitivity to imposed stretch and quantifying the Markstein number is essential for understanding the mechanisms underlying the frequency-dependent flame response and for enabling active control of instabilities in lean hydrogen-air flames.

The present work is based on a numerical dataset of two-dimensional (2D) lean hydrogen-air premixed laminar flames subjected to a range of imposed acoustic forcing conditions~\cite{young2025lean}. The main objective is to characterise the instability behaviour and to examine the correlation between flame speed and stretch rate at different equivalence ratios and acoustic forcing frequencies.

\section{Numerical Simulations\label{sec:simulation}} \addvspace{10pt}

The simulations are performed using the fully compressible direct numerical simulation (DNS) code \textsc{SENGA2}~\cite{Senga2}. The governing conservation equations are solved using a 10th-order central finite-difference scheme for spatial discretisation. Temperature-dependent thermophysical properties are evaluated from CHEMKIN polynomials, while molecular transport is described using a mixture-averaged model that includes Soret and Dufour effects, as well as the barodiffusion. The chemistry of hydrogen combustion is modelled by the skeletal mechanism of Burke \emph{et al.}~\cite{Burke2012}, which consists of 9 species and 23 reversible reactions.

\begin{figure}[h!]
\centering
\includegraphics[width=175pt]{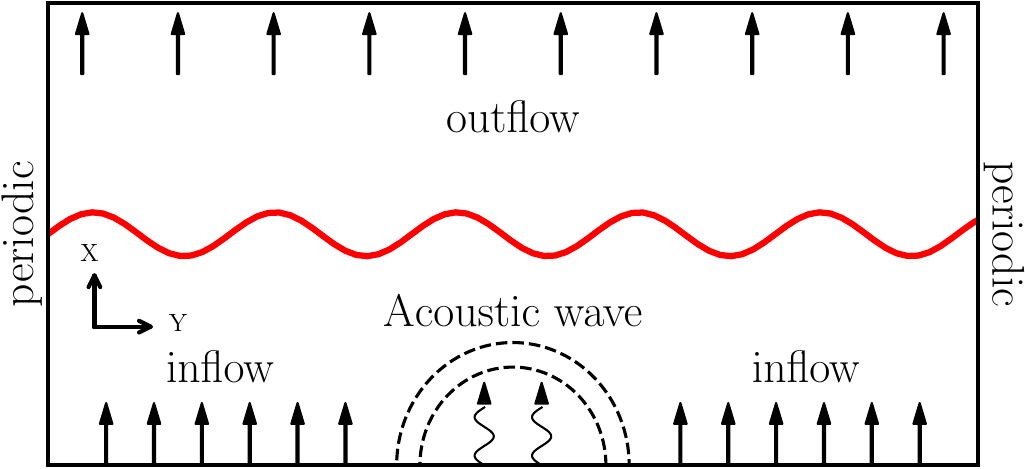}
\vspace{0.1 in}
\caption{Schematic of the 2D computational domain.}
\vspace{-5 pt}
\label{fig0}
\end{figure}

As shown schematically in Fig.~\ref{fig0}, two-dimensional simulations are performed for lean premixed H$_2$/air flames at equivalence ratios $\phi = 0.4$ and $\phi = 0.7$ under atmospheric conditions ($T_u = 300\,\mathrm{K}$, $p_{\mathrm{ref}} = 1\,\mathrm{bar}$). The flame is initialised from a fully developed, unstretched one-dimensional laminar flame profile positioned at the centre of the domain. Table \ref{tab:flame_acoustic_properties1} provides the details for the flame properties, including the thermal flame thickness $\delta_{\mathrm{th}}$, laminar flame speed $S_L$ and chemical timescale $\tau_{\mathrm{c}} = \delta_{\mathrm{th}}/S_L$. Acoustic forcing is introduced through a monopole-type source at the inflow boundary. In practice, this is implemented by imposing a sinusoidally varying inflow velocity over a small central section of the inlet using a top-hat spatial profile. The remainder of the inlet is maintained at a constant velocity equal to the laminar flame speed, $S_L$. %In this way, a localized time-periodic disturbance is generated at the inlet and subsequently propagates through the computational domain as an acoustic wave. 
The use of a fully compressible solver allows the acoustic pressure field and propagating pressure waves to be resolved directly, without any low-Mach-number approximation. Non-reflecting Navier-Stokes characteristic boundary conditions \cite{Lodato2008} are applied at the inflow and outflow boundaries, while periodic boundary conditions are used in the transverse direction. This configuration and inflow velocity profile is similar to the one used in the experiments of Searby \& Rochwerger \cite{Searby1991}.

In the present study, the acoustic excitation is imposed at $\mathrm{SPL}=130\,\mathrm{dB}$, with forcing frequencies ranging from 35 to 500\,kHz. 
%These frequencies are chosen such that the characteristic acoustic wavelength, and hence the associated pressure-gradient length scale, becomes comparable to the thermal flame thickness and the internal flame structure~\cite{mcintosh1999deflagration}. McIntosh~\cite{mcintosh1991high} demonstrated that the mass burning rate and generation of flame surface area are significantly affected by fast acoustic timescales, as well as the flame dynamics. Although the upper limit of this frequency range extends beyond that typically encountered in industrial applications, it is considered here because of its relevance to fundamental flame physics and its potential use in active-control strategies \cite{di2026experimental,poinsot1989initiation,lang1987active}. In particular, a 32.3 kHz ultrasonic field was employed recently for active control of ultra-lean hydrogen in \cite{di2026experimental}.
Although the upper limit of this frequency range extends beyond that typically encountered in industrial applications, it is considered here because of its relevance to the fundamental breakdown of the acoustically compact flame assumption. Many studies of combustion acoustics and combustion noise rely on such compact flames, in which the acoustic wavelength is much larger than the flame thickness. However, in a recent work \cite{KANG2026114785}, acoustically non-compact flames are shown to exist in the case of lean hydrogen. Therefore, this high frequency range is chosen such that the characteristic acoustic wavelength, and hence the associated pressure-gradient length scale, becomes comparable to the thermal flame thickness and the internal flame structure~\cite{mcintosh1999deflagration}. This allows to systematically examine the flame response across flames with varying degrees of compactness. In addition, high-frequency or ultrasonic forcing has also been explored as a potential strategy for
active combustion control. In particular, a 32.3 kHz ultrasonic field was employed recently for active control of ultra-lean hydrogen in \cite{di2026experimental}. The computational domain is chosen to accommodate multiple acoustic wavelengths and sufficiently large flame-front structures. For $\phi = 0.4$, the domain size is $L_x = 40\delta_{\mathrm{th}}$ and $L_y = 256\delta_{\mathrm{th}}$, while for $\phi = 0.7$ it is $L_x = 87\delta_{\mathrm{th}}$ and $L_y = 555\delta_{\mathrm{th}}$. All simulations use a uniform grid of $1280 \times 8192$ points. This ensures a minimum of 15 points within $\delta_{\mathrm{th}}$, as well as a minimum of
28 points resolving the acoustic wavelength $\lambda$, reflecting an adequate resolution for both the flame front \cite{Berger2022,Berger2022a,2024Kassar} and acoustic waves. Further details of the numerical methods, computational setup, and code validation are available in \cite{young2025lean}.

\begin{figure*}[ht!]
\centering
\makebox[\textwidth][c]{%
\includegraphics[width=0.5\textwidth]{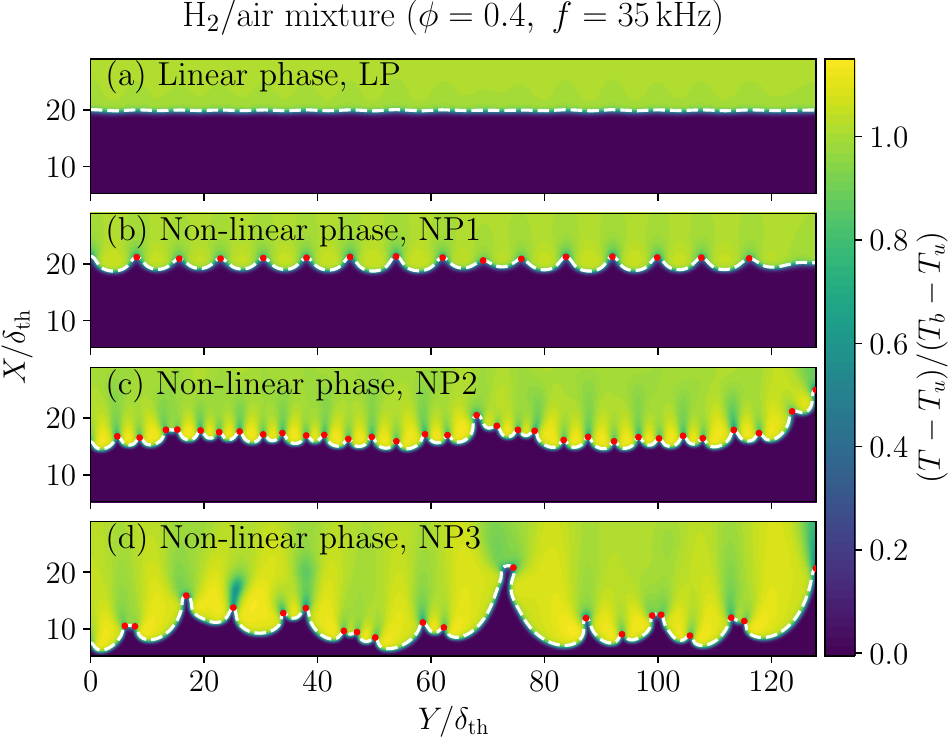}
\includegraphics[width=0.5\textwidth]{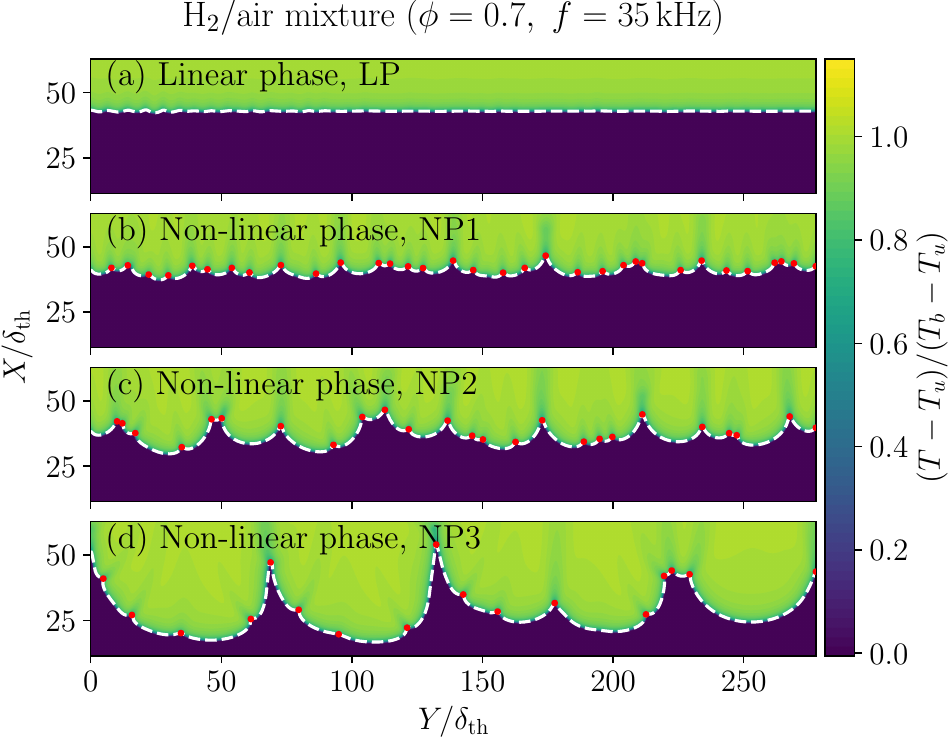}}
\vspace{-10 pt}
\caption{\footnotesize Temporal evolution of the normalized temperature field under acoustic forcing for H$_2$/air flames with $\phi = 0.4$ (left) and $\phi = 0.7$ (right) at $f = 35\,\mathrm{kHz}$ and $\text{SPL} = 130~\text{dB}$. Panels (a)–(d) correspond to time instants of $6.5\tau_{\mathrm{c}}$, $8.5\tau_{\mathrm{c}}$, $12.0\tau_{\mathrm{c}}$, and $16.3\tau_{\mathrm{c}}$ for $\phi = 0.4$, and $24.0\tau_{\mathrm{c}}$, $34.0\tau_{\mathrm{c}}$, $45.0\tau_{\mathrm{c}}$, and $65.0\tau_{\mathrm{c}}$ for $\phi = 0.7$, respectively. Due to the symmetry of the computational domain, only the left half of the domain is shown.}

\vspace{-10 pt}
\label{fig1}
\end{figure*}

\begin{table}[htbp]
{\small
\centering
\caption{Flame and acoustic properties. Acoustic wavelengths are calculated based on the reference sound speed of 340 m/s. The sound speed for the H$_2$/air mixture is 374 m/s at $\phi = 0.4$ and 392 m/s at $\phi = 0.7$.}
\label{tab:flame_acoustic_properties1}
{
\setlength{\tabcolsep}{2.5pt}
\begin{tabular}{lcccc}
\hline
\multicolumn{5}{c}{Flame properties} \\
\hline
$\phi$ & $S_L$ (m/s) & $\delta_{th}$ (mm) & $\tau_c$ (ms) & $\lambda_{peak}$ (mm) \\
\hline
0.4 & 0.17 & 0.76 & 4.47 & 4.22 \\
0.7 & 1.23 & 0.35 & 0.28 & 2.16 \\
\hline
\multicolumn{5}{c}{Acoustic properties} \\
\hline
$f$ (kHz) & 35 & 100 & 250 & 500 \\
\hline
$T_a$ (ms) & 0.029 & 0.010 & 0.004 & 0.002 \\
\hline
$\lambda_a$ (mm) & 9.71 & 3.40 & 1.36 & 0.68 \\
\hline
\end{tabular}
}}
\end{table}

\section{Results and discussion\label{sec:results}} \addvspace{10pt}
\subsection{Evolution of flame-front morphology
\label{subsec:subsection1}} \addvspace{10pt}

As the flame evolves, the imposed acoustic field interacts with the flame front and introduces wrinkles triggering the intrinsic instabilities of the flame. The subsequent evolution of flame-front morphology is governed by both the characteristics of the mixture and the acoustic forcing. These effects are examined in the following two subsections, which focus on the roles of equivalence ratio and acoustic forcing frequency, respectively.

\subsubsection{Effect of equivalence ratio}\addvspace{10pt}
\label{subsubsec:eqratio}
%\addvspace{10pt}
The evolution of acoustically forced lean premixed ${\mathrm{H_2}}$/air flames is first examined for equivalence ratios of $\phi = 0.4$ and $\phi = 0.7$ at a forcing frequency of $f = 35\,\mathrm{kHz}$. Figure~\ref{fig1} shows the spatial distribution of the normalized temperature $\Theta = (T - T_u)/(T_b - T_u)$, where $T_u$ and $T_b$ are the unburned and burned temperature of the mixtures, respectively. In the present cases, the burned temperature is approximately $T_{b} \approx 1423\,\mathrm{K}$ for $\phi = 0.4$ and $T_{b} \approx 2016\,\mathrm{K}$ for $\phi = 0.7$. Due to the symmetry of the computational domain, only the left half of the domain is shown in Fig. \ref{fig1}.

%\vspace{-5 pt}
\begin{figure}[h!]
\centering
\includegraphics[width=180pt]{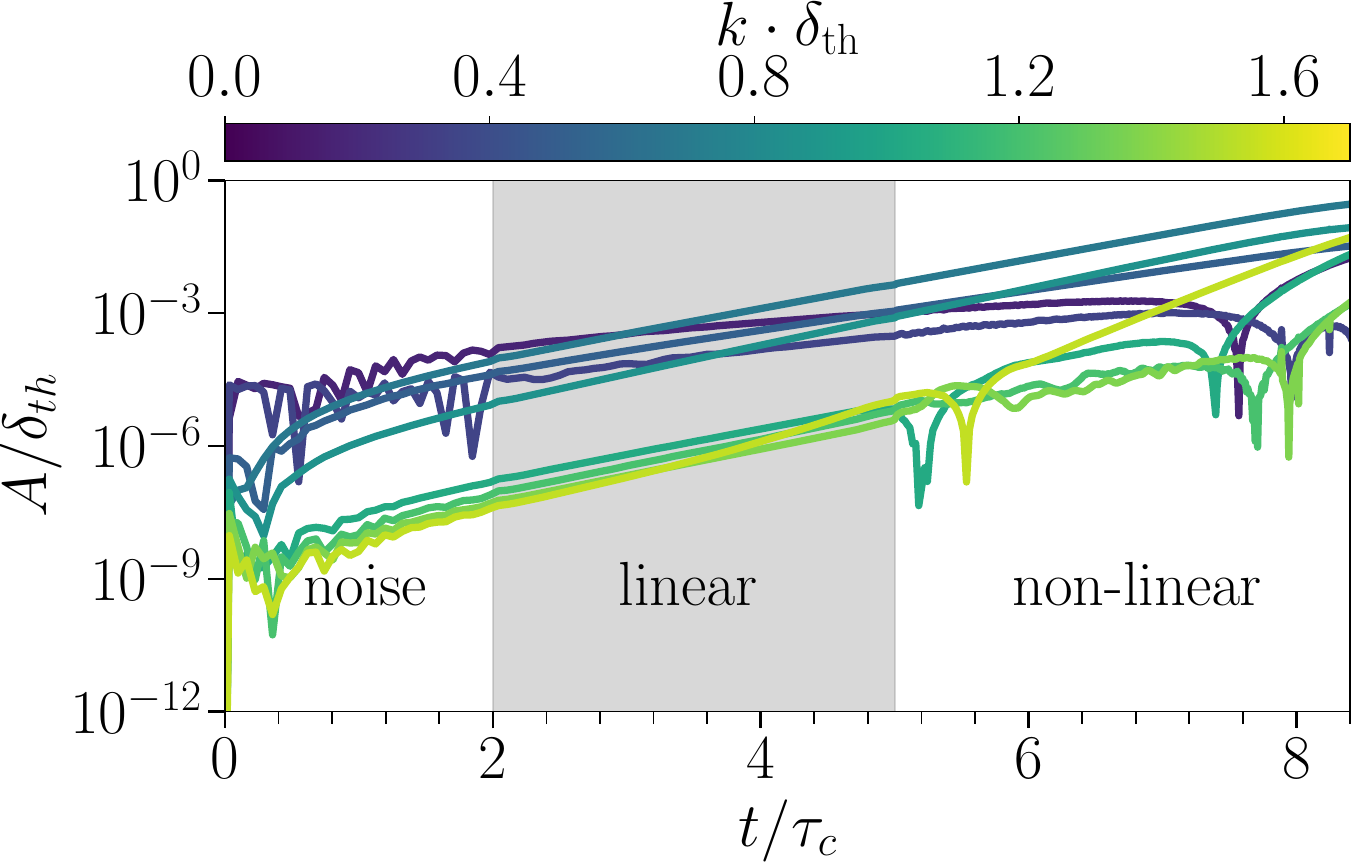}
\vspace{2 pt}
\caption{\footnotesize Amplitude evolution of different Fourier modes in a logarithmic scale for H$_2$/air mixture with $\phi = 0.4$, $f = 35~\text{kHz}$ and $\text{SPL} = 130~\text{dB}$.}
\vspace{-5 pt}
\label{fig2}
\end{figure}

For both equivalence ratios, the flame initially experiences weak perturbations induced by the acoustic forcing. During this stage, the perturbations remain small and the flame front evolution is well described by a linear growth phase (i.e. LP in Fig.~\ref{fig1}). To quantify this behavior, the temporal evolution of the Fourier mode amplitudes $A(t)$ associated with different wavenumbers $k$ is analysed. As illustrated in Fig.~\ref{fig2}, the amplitude of each perturbation mode exhibits exponential growth in the linear phase, following $A(t) = A_0 \exp(\omega t)$, where $\omega$ is the growth rate of the perturbation. The growth rate is obtained from $\omega = d[\ln(A(t))]/dt$, i.e. the slope of $\ln(A(t))$ during the linear stage. By evaluating $\omega$ over a range of wavenumbers, the dispersion relation $\omega(k)$ can be constructed following the standard linear stability analysis of premixed flames. A systematic analysis of dispersion relations for different equivalence ratios can be found in our recent work ~\cite{young2025lean}. Based on Fig.~\ref{fig2}, the time intervals corresponding to the linear and non-linear phases can be approximately identified. The flame patterns encountered in each regime are illustrated in Fig.~\ref{fig1}.

\begin{figure}[h!]
\centering
\includegraphics[width=180pt, trim=0 0 0 0, clip]{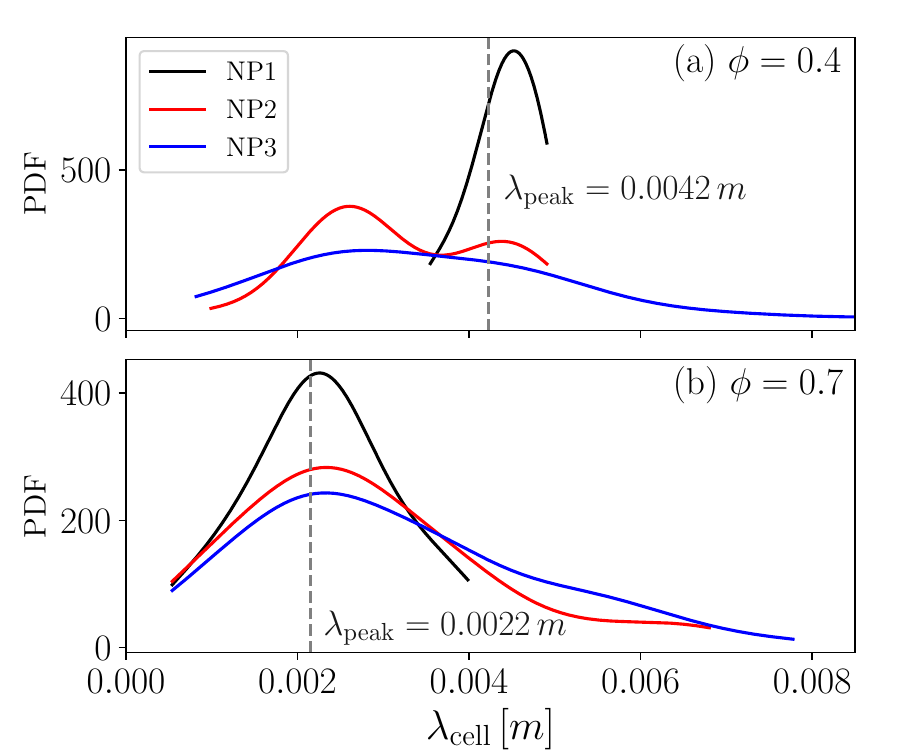}
\vspace{2 pt}
\caption{\footnotesize Probability distributions of flame-front length scales during the non-linear phase for H$_2$/air mixtures at $f = 35\,\mathrm{kHz}$ and $\mathrm{SPL} = 130\,\mathrm{dB}$: (a) $\phi = 0.4$, at $8.5\tau_{\mathrm{c}}$ (NP1), $12.0\tau_{\mathrm{c}}$ (NP2), and $16.3\tau_{\mathrm{c}}$ (NP3); and (b) $\phi = 0.7$, at $34.0\tau_{\mathrm{c}}$ (NP1), $45.0\tau_{\mathrm{c}}$ (NP2), and $65.0\tau_{\mathrm{c}}$ (NP3). The dashed vertical line marks $\lambda_{\mathrm{peak}}$, the most unstable wavelength predicted by linear stability analysis \cite{young2025lean}.}
\vspace{-10 pt}
\label{fig3}
\end{figure}

To assess the structure of the flame front, the flame position is identified using a progress variable. The progress variable is defined as $c = (Y_{\mathrm{H_2}} - Y_{\mathrm{H_2},u})/(Y_{\mathrm{H_2},b} - Y_{\mathrm{H_2},u})$, where $Y_{\mathrm{H_2}}$ is the hydrogen mass fraction and the subscripts $u$ and $b$ denote the unburned and burned states, respectively. The instantaneous flame front is identified by the iso-contour corresponding to the location of maximum heat release rate in an unstretched, freely propagating laminar flame. This is indicated by the white dashed lines in Fig.~\ref{fig1}. The characteristic length scales of the flame-front corrugations are quantified from the extracted iso-lines. In particular, the cell size, $\lambda_{\mathrm{cell}}$, is defined as the distance between two consecutive concave cusps along the flame front. These cusps are identified as negative extrema of the local curvature distribution and are marked by red dots in Fig.~\ref{fig1}. Assuming that the flame segment between two neighbouring cusps is approximately semi-circular, the cell size is estimated as $\lambda_{\mathrm{cell}}={2l^*}/{\pi}$, where $l^*$ denotes the arc length between the two cusps \cite{berger2019characteristic,berger2022intrinsic}. The resulting distribution of $\lambda_{\mathrm{cell}}$ is shown in Fig.~\ref{fig3}, providing a quantitative measure of the flame cell length scale.

%\vspace{-7 pt}

\begin{figure*}[ht!]
\centering
\vspace{-0.4 in}
\includegraphics[width=1.05\textwidth]{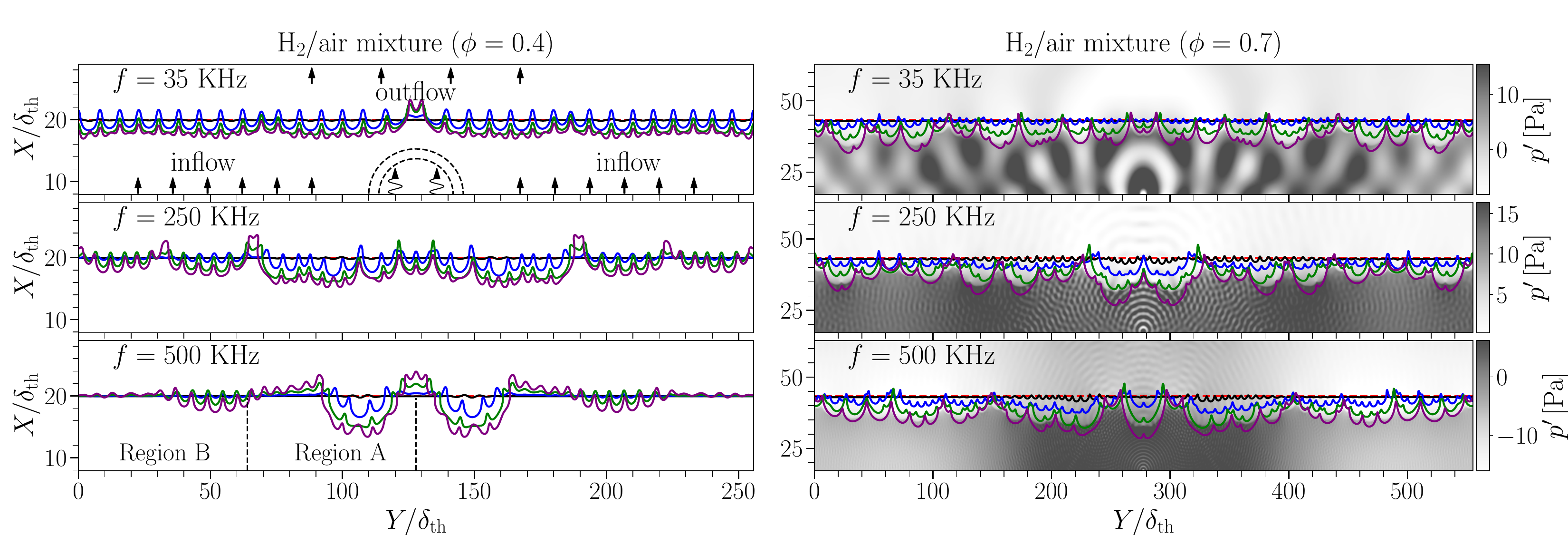}
\vspace{-10 pt}
\caption{\footnotesize Temporal evolution of the flame front under acoustic forcing for H$_2$/air flames with (a) $\phi = 0.4$ and (b) $\phi = 0.7$. The black, blue, green, and purple lines correspond to time instants of 6.7$\tau_{\mathrm{c}}$, 9.0$\tau_{\mathrm{c}}$, 10.0$\tau_{\mathrm{c}}$, 10.7$\tau_{\mathrm{c}}$ for $\phi = 0.4$, and 20.0$\tau_{\mathrm{c}}$, 28.0$\tau_{\mathrm{c}}$, 36.0$\tau_{\mathrm{c}}$, 42.0$\tau_{\mathrm{c}}$ for $\phi = 0.7$, respectively. The inflow frequencies are $f$=35~kHz, 250~kHz, and 500~kHz from top to bottom.}
\vspace{-15 pt}
\label{fig4}
\end{figure*}

As shown in Fig.~\ref{fig1}, for the leaner mixture ($\phi = 0.4$), the flame dynamics are strongly affected by thermodiffusive instability due to the low effective Lewis number ($Le_{eff}$). The flame-front evolution follows a characteristic uniform (NP1), split (NP2), merge (NP3) sequence. In the early non-linear stage (NP1), a relatively regular cellular structure develops along the flame front. This gives rise to a pronounced peak at a characteristic wavelength in the PDF shown in Fig.~\ref{fig3}(a). This dominant cell size is very close to the most unstable wavelength, $\lambda_{\mathrm{peak}}$ predicted by \cite{young2025lean}. This suggests that the initial non-linear cellular structures follow from the linear growth phase where the most unstable wave numbers tend to dominate, consistent with \cite{berger2019characteristic,lulic2023polyhedral}. In \cite{young2025lean}, it is shown that the peak wavelength is insensitive to the imposed acoustic forcing frequency. The normalised peak wavenumber $\bar{k}_{\mathrm{peak}} = k_{\mathrm{peak}}\delta_{\mathrm{th}}$ was reported as $\bar{k}_{\mathrm{peak}} = 1.13$ (i.e. $\lambda_{\mathrm{peak}} = 4.22\,$mm) for $\phi = 0.4$, and $\bar{k}_{\mathrm{peak}} = 1.02$ (i.e. $\lambda_{\mathrm{peak}} = 2.16\,$mm) for $\phi = 0.7$, as shown in Table \ref{tab:flame_acoustic_properties1}. As the unstable flame develops further (NP2), the initially regular cells split into smaller segments, and the corresponding PDF broadens towards smaller wavelengths. At the same time, a second peak emerges at a smaller wavelength, indicating the formation of finer-scale flame-front corrugations. At later times (NP3), neighbouring cells periodically undergo a merging process and give rise to larger flame structures, including flame fingers characterized by strongly curved cusps at their tips. Correspondingly, the PDF broadens further towards larger wavelengths, showing the growth of large-scale structures that contain embedded smaller-scale flame-front corrugations. During the strongly unstable non-linear stage, NP1-NP3, super-adiabatic temperature regions are also observed, where the local temperature exceeds the adiabatic flame temperature. These regions typically appear in the wake of convex flame segments and are associated with locally enhanced reaction rates induced by preferential diffusion effects. This is in agreement with the established picture of thermodiffusively unstable lean hydrogen flames \cite{berger2019characteristic}.

In contrast, the $\phi = 0.7$ mixture exhibits weaker instability, as the larger $Le_{eff}$ reduces the influence of thermodiffusive effects. In this case, the flame dynamics are more strongly influenced by the hydrodynamic (Darrieus–Landau, DL) instability associated with gas expansion across the flame front. Although cellular structures still develop, the flame front remains comparatively smooth during the early non-linear evolution. As shown in Fig.~\ref{fig3}(b), the PDF in the initial non-linear stage (NP1) is also centred around the peak wavelength, $\lambda_{\mathrm{peak}} = 2.2\,$mm, predicted by the linear stability analysis \cite{young2025lean}, and the initial cell size is smaller than that in the $\phi = 0.4$ case. As a result, the flame does not exhibit a clearly identifiable splitting stage. Instead, the subsequent evolution (NP2-NP3) is dominated primarily by cell merging and large-scale wrinkle growth. Consistently, the PDFs broaden progressively with time and shift towards larger wavelengths, reflecting the gradual development of larger flame-front structures. This behaviour is consistent with the weaker thermodiffusive effect and the relatively greater role of DL-driven corrugation in the $\phi = 0.7$ flame \cite{berger2019characteristic,berger2022intrinsic}. 

\subsubsection{Effect of acoustic forcing frequency
\label{subsubsec:frequency}}
\addvspace{10pt}

Figure~\ref{fig4} shows the temporal evolution of flame-front corrugations during the transition from the linear to the non-linear regime (NP1-NP2) under acoustic forcing for both equivalence ratios $\phi = 0.4$ and 0.7. Results are presented for forcing frequencies of 35, 250, and 500~kHz, in order of increasing frequency. A clear frequency dependence is evident in the non-linear flame response. At lower forcing frequencies, the flame-front corrugations remain relatively uniform along the flame surface. As the forcing frequency increases the flame front becomes progressively modulated and develops envelope-like structures, resembling a carrier-envelope pattern.

\begin{figure}[h!]
\centering

\includegraphics[width=190pt, trim=0 0 0 90, clip]{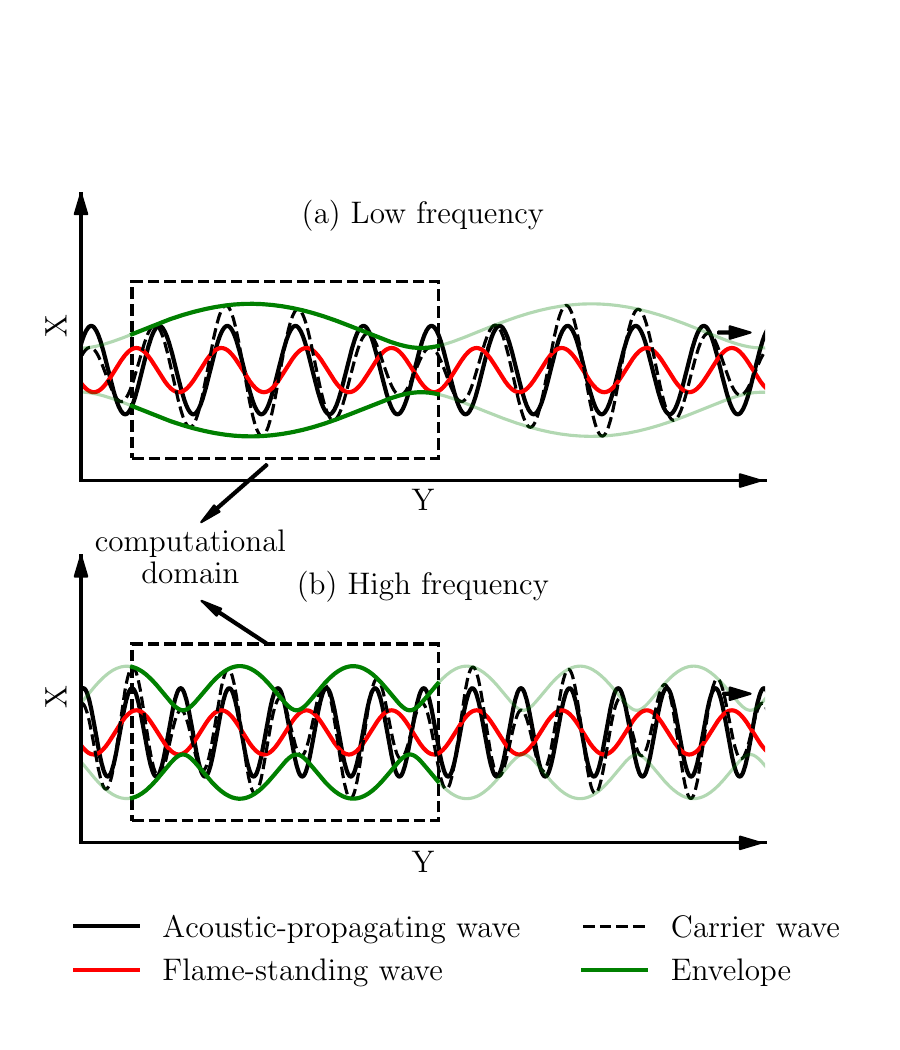}
\vspace{-20 pt}
\caption{\footnotesize Schematic of flame-acoustic interactions.}
%\vspace{-15 pt}
\label{fig5}
\end{figure}

A possible interpretation of this behaviour is illustrated schematically in Fig.~\ref{fig5}. As a first-order conceptual approximation, the flame-front displacement is represented as the superposition of an intrinsic standing cellular mode, $X_1(Y) = \sin(k_1 Y + c_1)$, and an acoustically forced travelling mode, $X_2(Y,t) = \sin(k_2 Y + \Omega t + c_2)$. Here, $X_1$ represents the intrinsic flame-front corrugation associated with the cellular instability and $X_2$ denotes the perturbation introduced by the imposed acoustic forcing. The wavenumber $k_1$ is taken as the most unstable mode identified from the dispersion relation, i.e.\ $k_1 \approx k_{\mathrm{peak}}$, whereas $k_2$ is the wavenumber of the acoustically forced travelling wave and $\Omega = 2\pi f$ is the angular frequency corresponding to the imposed forcing frequency $f$. The instantaneous flame-front shape is therefore approximated as $X(Y,t) = X_1(Y) + X_2(Y,t)$. At a fixed instant $t=t_0$, this becomes $X_0(Y,t_0)=\sin(k_1 Y+c_1)+\sin(k_2 Y+\phi_0)$, where $\phi_0=\Omega t_0+c_2$. Using the trigonometric identity for the sum of two sine waves, the flame-front shape can be rewritten as
\begin{equation}
\begin{aligned}
X_0(Y,t_0)
&= 2 \sin\!\left[\frac{(k_1+k_2)Y+c_1+\phi_0}{2}\right] \\
&\quad \cdot \cos\!\left[\frac{(k_1-k_2)Y+c_1-\phi_0}{2}\right].
\end{aligned}
\end{equation}
This expression shows that the flame front consists of a rapidly varying carrier wave modulated by a slowly varying envelope. The corresponding envelope amplitude is
\begin{equation}
A_{\mathrm{env}}(Y,t_0)
=
2\left|
\cos\!\left[\frac{(k_1-k_2)Y+c_1-\phi_0}{2}\right]
\right|,
\end{equation}
with a spatial modulation wavenumber of $|k_2-k_1|/2$. The corresponding envelope wavelength is $\lambda_{\mathrm{env}}={4\pi}/{|k_2-k_1|}$. When the forcing frequency is low, e.g. $f$=35~kHz, the mismatch between the intrinsic and forced spatial modes remains small, giving rise to a long envelope wavelength and hence nearly uniform flame wrinkling over the computational domain. As the forcing frequency increases, e.g. $f$=500~kHz, the mismatch becomes larger, shortening the envelope wavelength and producing increasingly non-uniform flame-front structures with pronounced envelope-like patterns within the finite observation window.

%\vspace{-7 pt}
\begin{figure}[h!]
\centering
\includegraphics[width=180pt, trim=0 0 0 0, clip]{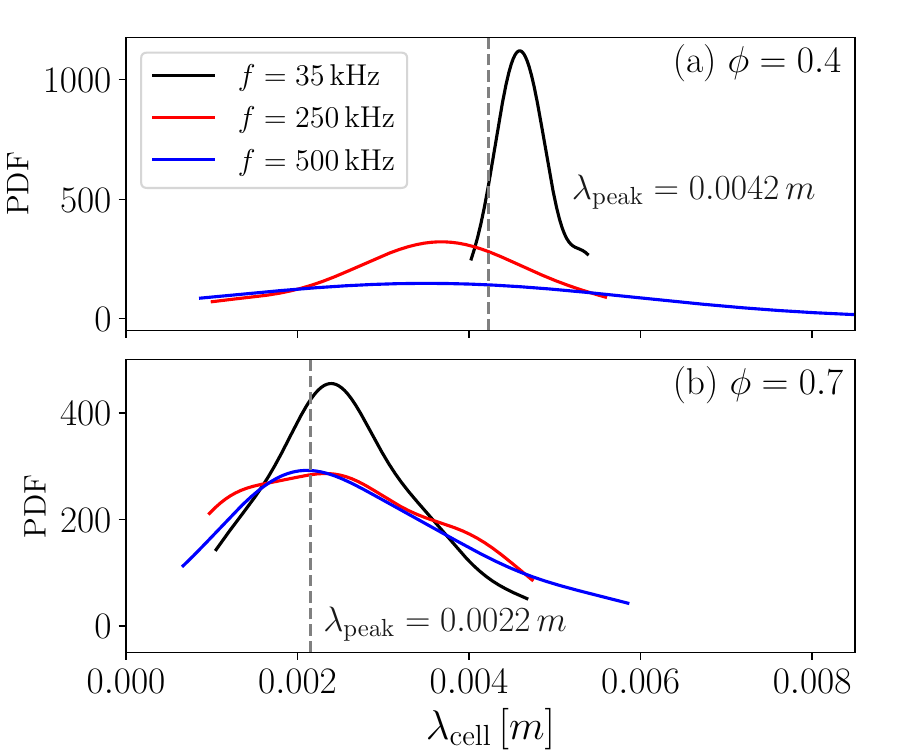}
\vspace{2 pt}
\caption{\footnotesize Probability distributions of flame-front length scales during the non-linear phase for H$_2$/air mixtures with (a) $\phi = 0.4$ at $10\tau_{\mathrm{c}}$ and (b) $\phi = 0.7$ at $36\tau_{\mathrm{c}}$, for acoustic forcing frequencies of $f = 35$, $250$, and $500\,\mathrm{kHz}$ at $\mathrm{SPL} = 130\,\mathrm{dB}$. The dashed vertical line marks $\lambda_{\mathrm{peak}}$, the most unstable wavelength predicted by linear stability analysis \cite{young2025lean}.}
%\vspace{-5 pt}
\label{fig6}
\end{figure}

While the simplified superposition model provides a first-order interpretation of the flame response, the actual influence of acoustic perturbations on the flame front is considerably more complex. In particular, the flame front cannot be expected to retain a fixed characteristic mode throughout its evolution due to the persistent perturbation imposed by the acoustic forcing. In practice, pressure fluctuations modify the local flow field in the vicinity of the flame, thereby altering the strain rate and curvature experienced by the flame front \cite{Searby1991}. The flame responds dynamically to these perturbations. The nature of this response depends strongly on the Markstein number, which governs the sensitivity of the local burning velocity to stretch and reflects the relative importance of thermal and mass diffusion \cite{markstein1964}. A comparison of the flame structures in Fig.~\ref{fig4} shows that the $\phi = 0.4$ case exhibits stronger and more localized instability near the centre of the domain (Region A) at $f = 500\,\mathrm{kHz}$, whereas the $\phi = 0.7$ case displays a more spatially uniform response along the flame front. This difference can be attributed to the smaller Markstein number of the $\phi = 0.4$ ${\mathrm{H_2}}$/air mixture, with intrinsic flame instabilities resulting in rapid growth of corrugations in the central region where the flame is first perturbed. Once established, this strong local response weakens the transmission of the acoustic disturbance towards the lateral regions (Region B in Fig.~\ref{fig4}), which remain comparatively less affected. In contrast, the $\phi = 0.7$ mixture with its weaker thermodiffusive instability, allows the acoustic perturbation to influence the flame front more uniformly across the domain.

The effect of forcing frequency on the non-linear flame behaviour is quantitatively reflected in the PDFs of $\lambda_{cell}$ shown in Fig.~\ref{fig6}. For $\phi = 0.4$, the PDF broadens markedly with increasing forcing frequency, indicating the emergence of a wider range of flame-front corrugation scales. This broadening reflects the combined effect of the carrier-envelope modulation induced by acoustic forcing and the strong thermodiffusive instability of the leaner mixture. In particular, the extension of the PDF towards larger $\lambda_{\mathrm{cell}}$ corresponds to the formation of large-scale structures in Region A of Fig.~\ref{fig4}, while the overall broad distribution reflects the coexistence of these large central structures with smaller, weakened cells in the lateral regions (i.e. Region B). For $\phi = 0.7$, a similar frequency-dependent broadening of the PDF is also observed, but it is less pronounced. This can be attributed to the weaker thermodiffusive instability of the $\phi = 0.7$ flame, which leads to less distinct scale separation and, consequently, a comparatively narrower distribution than in the $\phi = 0.4$ case. 

Another limitation of the proposed conceptual framework is that it does not explicitly account for the effect of acoustic forcing amplitude. The framework characterizes flame front evolution as the superposition of two modes: an intrinsic standing cellular mode associated with thermo-diffusive instability, and an acoustically forced travelling mode. A prerequisite for this interpretation is that the flame be sufficiently perturbed by the acoustic wave to allow the intrinsic cellular mode to develop. However, as shown in the supplementary material, for the $\phi$ = 0.7 flame, differential diffusion effects are weaker. Under the lower forcing amplitude of 110 dB, the acoustic perturbation is not strong enough to penetrate the lateral region and trigger a uniform, well-developed intrinsic cellular structure. As a result, the acoustic wave perturbs only the central region of the flame front, leading to a more centre-localised pattern. This observation suggests that the current framework has limitations when the forcing amplitude is too weak to excite the intrinsic cellular mode.

\subsection{Correlation between flame speed and stretch \label{subsec:subsection2}} \addvspace{5pt}

\begin{figure*}[h!]
\centering
\hspace{-30pt} 
\makebox[\textwidth][c]{
\includegraphics[width=0.5\textwidth]{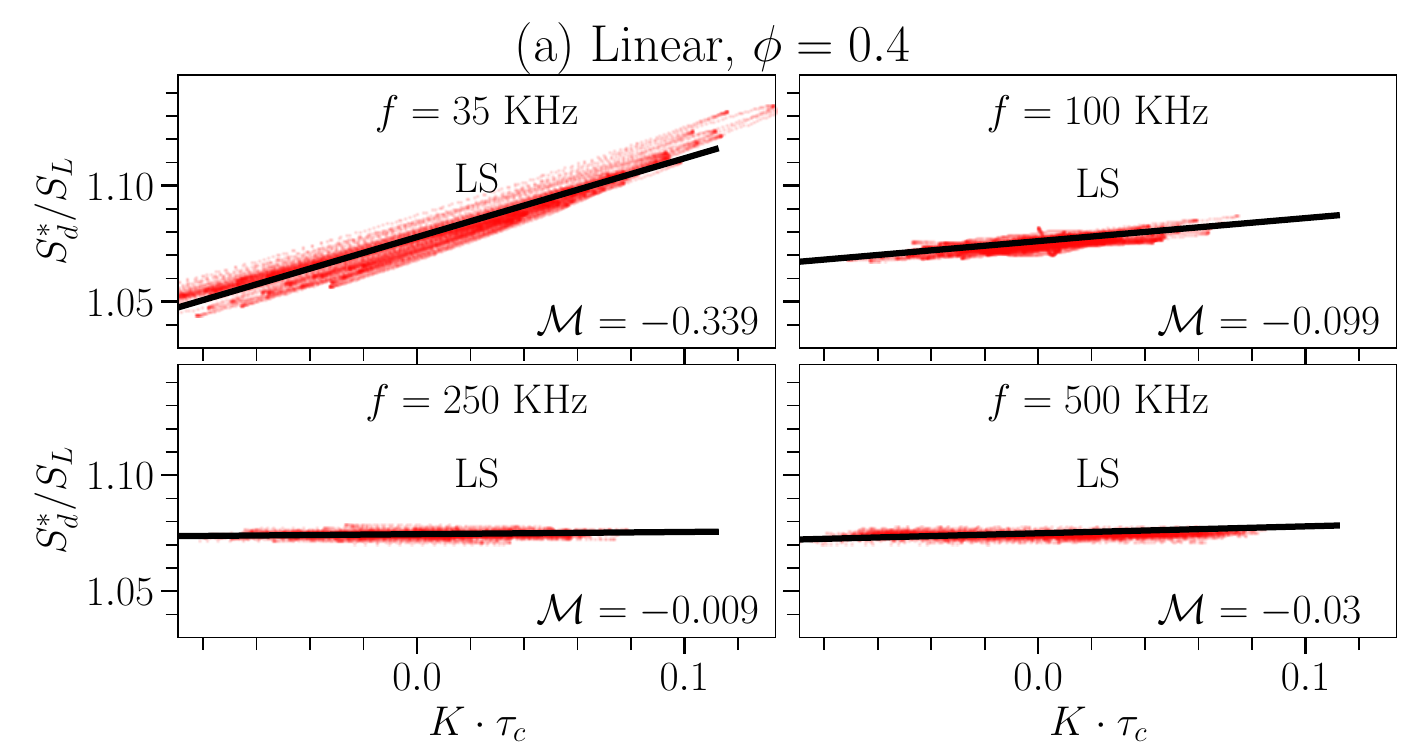}
\includegraphics[width=0.5\textwidth]{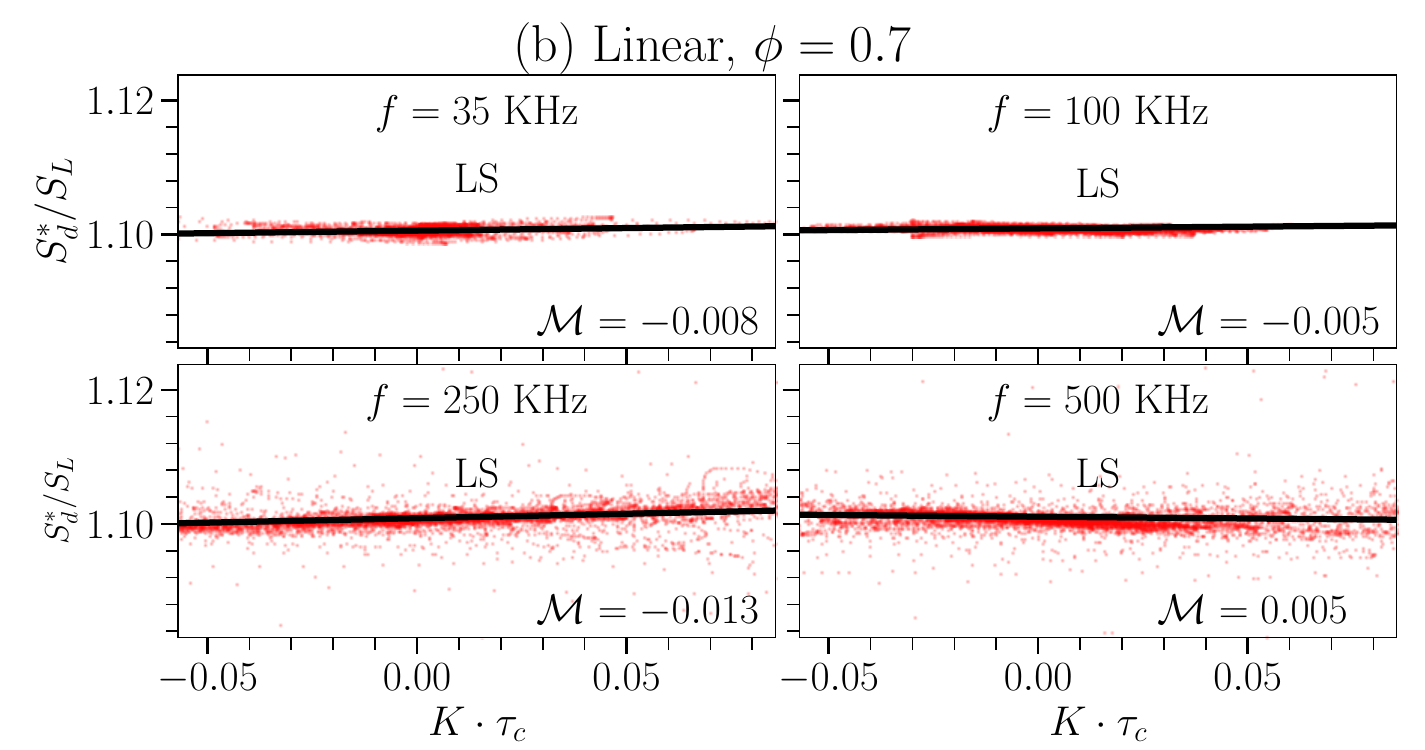}
}

\hspace{-30pt} 
\makebox[\textwidth][c]{
\includegraphics[width=0.5\textwidth]{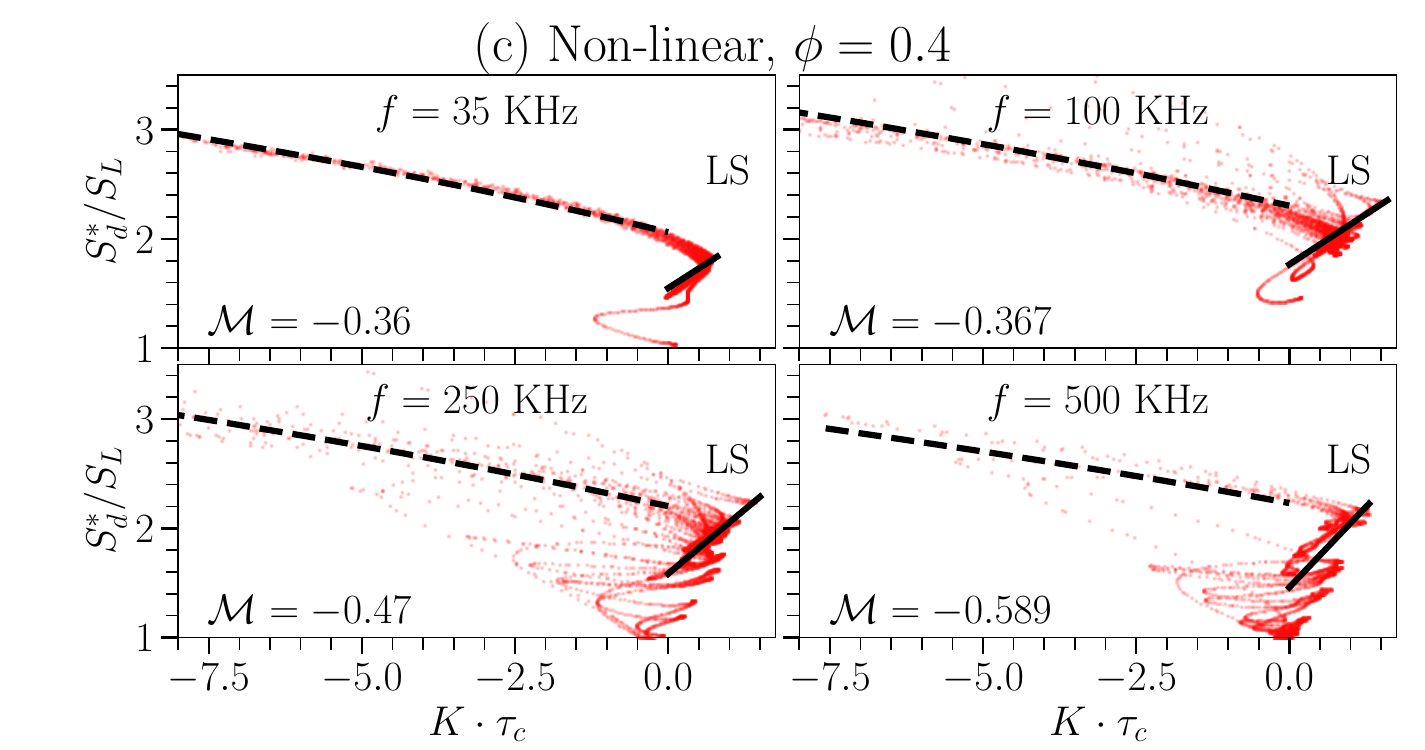}
\includegraphics[width=0.5\textwidth]{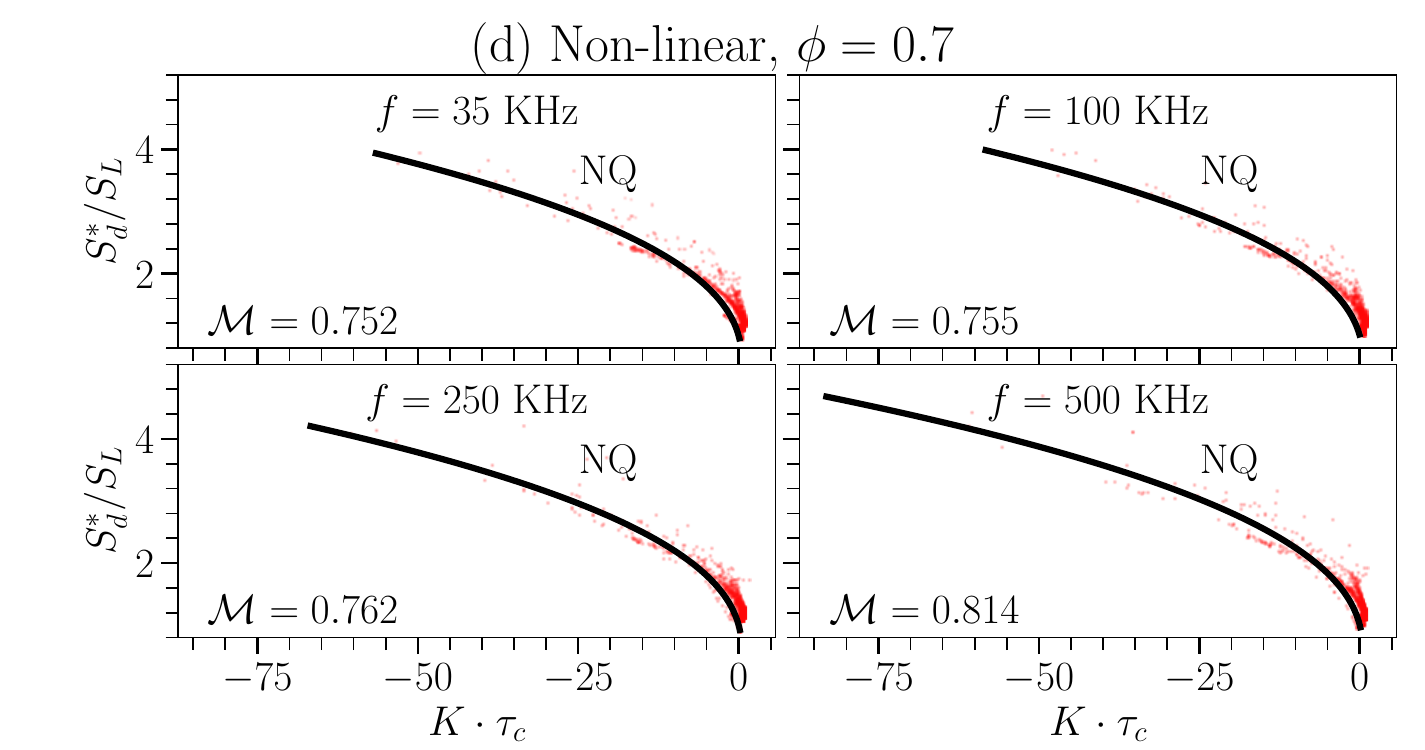}
}

\caption{\footnotesize Correlation of normalised density weighted displacement speed, ${S_d}^*/S_L$, with the normalised total stretch rate, $K\cdot\tau_c$, for (a-b) linear phase, $\phi = 0.4$ and $\phi = 0.7$; (c-d) non-linear phase, $\phi = 0.4$ at $10\tau_c$ and $\phi = 0.7$ at $36\tau_c$ for different frequencies $f$=35~kHz, 100~kHz, 250~kHz, and 500~kHz. LS fits are performed for (a–c); and NQ fit is performed for (d).}
\vspace{-12pt}
\label{fig7}
\end{figure*}

In order to gain further insight into the flame response to the imposed acoustic perturbation, the correlation between flame speed and stretch is examined. In a weakly stretched flame, the dependence of flame speed on flame stretch can be described either by a linear stretch (LS)~\cite{wu1985determination}, $S^*_d=S_L-\mathcal{M}K\delta_{th}$ or by a quasi-steady non-linear (NQ) extrapolation relation~\cite{kelley2009nonlinear}, $(S^*_d/S_L)^2\mathrm{ln}(S^*_d/S_L)^2=-\mathcal{M}K\delta_{th}/{S_L}$ with the help of Markstein number $\mathcal{M}$. Here, $S_d$ is the flame displacement speed defined as $S_d=\frac{\dot{\omega}_c+\nabla\cdot\left(\rho D_c\nabla c\right)}{\rho\left|\nabla c\right|}$ and $S^*_d=\rho S_d/\rho_u$ is the density-weighted displacement speed, where $\rho$ is the density, $\dot{\omega}_c$ is the chemical reaction rate of $c$ and $D_c$ is the diffusivity of $c$. The total stretch rate is given by $K= a_T+2S_d\kappa_m$ \cite{Candel1990}, where $a_T=(\delta_{ij}-n_in_j)\partial u_i/\partial x_j$ is the tangential strain rate, ${\bf n}=-\nabla c/|\nabla c|$ is the flame normal vector, and $\kappa_m=0.5\nabla\cdot{\bf n}$ is the arithmetic mean of the two principal flame curvatures. Figure~\ref{fig7} shows the variation of normalized $S^*_d$ with total stretch rate, $K$, for the two equivalence ratios under different acoustic forcing frequencies. The data are extracted at the leading edge of the flame, i.e.,
at $c$ = 0.1, since it is commonly adopted in turbulent combustion models including flame surface density (FSD) and scalar dissipation rate (SDR) models \cite{HAKBERG1985225}. Furthermore, turbulent flame speed models are often based on Kolmogorov, Petrovsky, and Piskunov (KPP) theorem which relies on the assumption of flame leading edge \cite{Poinsot2001Theoretical}. Similar trends are also observed at the location of peak heat release rate, as shown in the supplementary material.

The scatter plots in Figs.~\ref{fig7}(a) and \ref{fig7}(b) reveal an approximately linear relationship between $S^*_d$ and $K$, indicating that the LS model can be used to estimate the Markstein length during the linear phase of flame evolution. For the $\phi = 0.4$ flame, a positive correlation between $S^*_d$ and $K$ is observed at the lowest forcing frequency (35~kHz), corresponding to a negative Markstein length. This is consistent with previous findings for lean $\mathrm{H_2}$/air flames~\cite{im2000effects,chen2021effects}, which reflects the strong preferential-diffusion effect. As the forcing frequency increases, the Markstein length moves progressively towards zero. This trend suggests that, above a certain frequency threshold, the flame response becomes increasingly insensitive to stretch, consistent with the findings of Joulin~\cite{Joulin1994}. %During the linear phase, the flame front remains nearly flat (see Fig.\ref{fig1}), so its response is governed primarily by the imposed strain rate perturbation.
In the linear regime, the flame remains nearly flat and unwrinkled (i.e., zero curvature) and the flame is primarily perturbed by the strain rate. At high forcing frequencies, the characteristic timescale of the imposed flow becomes too short relative to the flame transit time. As a result, the influence of strain rate weakens, and the flame responds more like a unity Lewis number flame~\cite{Joulin1994}. In the case of $\phi = 0.7$ flame, the Markstein length remains close to zero throughout the linear regime over the entire forcing-frequency range considered here. This suggests that, the flame response is only weakly affected by the imposed perturbation for $\phi = 0.7$, with stretch playing a minor role in local flame propagation. This is expected for flames with $Le_{eff}$ closer to unity. Moreover, it is noteworthy that throughout the entire range of high forcing frequencies considered in this study, the characteristic timescale of the imposed flow (i.e. the period of acoustic wave, $T_a$) is consistently shorter than the flame transit time (i.e. $\tau_c$) as shown in Table \ref{tab:flame_acoustic_properties1}. As a consequence, the influence of strain rate on the flame is weak, and the normalized displacement speed in the linear regime takes values only marginally above unity.

\begin{figure}[h!]
\centering
\hspace{-30pt}
\includegraphics[width=180pt]{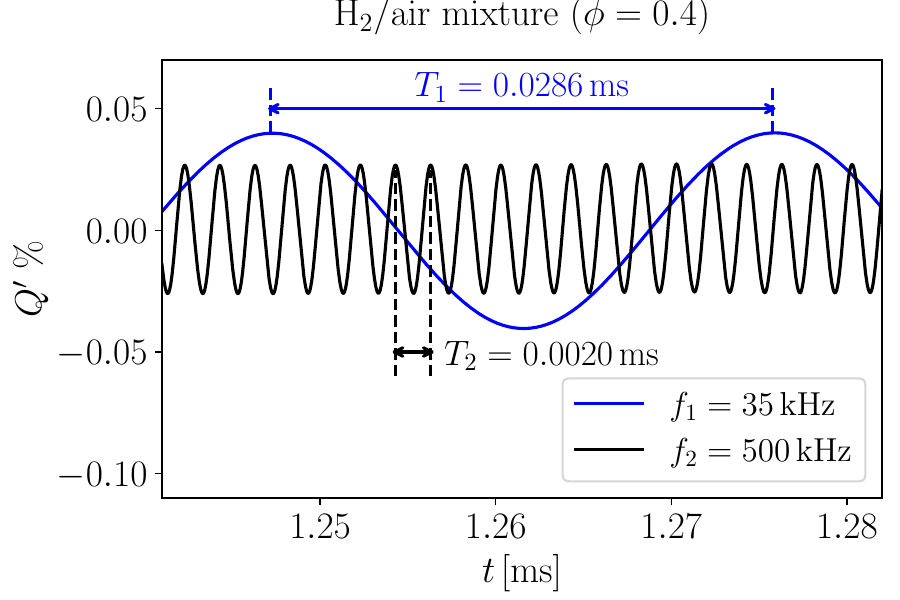}
\caption{\footnotesize Relative change in the integrated heat release for H$_2$/air flames with $\phi = 0.4$ and $f$=35~kHz, 500~kHz.}
%\vspace{-21 pt}
\label{fig8}
\end{figure}

Furthermore, it is noted that the present simulations are performed using a fully compressible solver, which resolves propagating pressure waves. As a result, both the hydrodynamic effects of pressure and the acoustic effects of wave propagation are inherently taken into account. To isolate the effect of acoustic perturbations on combustion dynamics, an additional set of one-dimensional simulations is performed for a premixed laminar hydrogen flame subjected to acoustic forcing. Consistent with the two-dimensional configuration, the same time dependent, sinusoidally varying inflow velocity as that in the corresponding 2D simulations is imposed at the inlet. Non-reflecting Navier-Stokes characteristic boundary conditions are applied at both the inflow and outflow boundaries. The integrated heat release rate of the acoustically perturbed flame, $Q$, is then compared with that of the unperturbed one-dimensional reference flame, $Q_0$, through $Q'(t)={\int (Q-Q_0)\,dx}/{\int Q_0\,dx}$. The time histories of $Q'$ are shown in Fig.~\ref{fig8} for acoustic forcing frequencies of $f=35$ and $500\,\mathrm{kHz}$ at $\phi=0.4$ and SPL=130 dB. In both cases, the flame response oscillates at the same frequency as the imposed acoustic forcing. Accordingly, the periods of the heat-release response are $T_1 = 1/f_1 = 0.0286\,\mathrm{ms}$ for $f_1=35\,\mathrm{kHz}$ and $T_2 = 1/f_2 = 0.0020\,\mathrm{ms}$ for $f_2=500\,\mathrm{kHz}$. Similar frequency locking has also been reported for methane/air flames in \cite{qureshi2013behaviour}. For the forcing frequencies considered here, the characteristic disturbance length scale, $c_0/(2f)$, ranges from $5.3\,\mathrm{mm}$ to $0.374\,\mathrm{mm}$, where $c_0$ is the speed of sound. At the higher forcing frequency, this length scale becomes comparable to the flame thickness ($0.76\,\mathrm{mm}$). Under such conditions, pressure oscillations can penetrate the internal flame structure. Their influence is exerted through the associated pressure gradients, which induce preferential acceleration of the lighter species, as well as through instantaneous local pressure variations within the flame, which can modify chemical kinetics and transport processes in the reaction zone \cite{mcintosh1999deflagration,mcintosh1991high}. However, as shown in Fig.~\ref{fig8}, the relative variations in heat release remain very small, and the response amplitude decreases as the forcing frequency increases. This trend can be interpreted in terms of the acoustic time-scale parameter proposed by McIntosh~\cite{mcintosh1999deflagration}. For harmonic forcing, this parameter can be written as $\tau_a = {\tau_{c}}/{\tau_{acoustic}}$= ${2 f \delta_{\rm th}}/{S_L}$. %which shows an explicit dependence on forcing frequency. 
As the frequency increases, the acoustic time scale becomes shorter, leaving less time within each cycle for the flame to adjust diffusively and chemically to the imposed oscillation. As a result, although the flame continues to respond at the forcing frequency, the integrated heat release rate becomes progressively less sensitive to acoustic perturbations at high frequencies. It is noted that though the relative variations in integrated heat release remain small in the one-dimensional simulations, these results remain valuable because they demonstrate that point-wise pressure variations and pressure-gradient-induced differential diffusion of species provide a physical mechanism through which acoustic forcing can modify flame structure. In the planar 1D configuration, the response is limited by the absence of curvature, transverse diffusion, and multidimensional hydrodynamic strain. In higher-dimensional flames, however, this mechanism may become more important through its coupling with these effects. This will potentially modify the flame morphology and leading to a stronger response to acoustic perturbation. More details on the numerical setup and results for the one-dimensional acoustically perturbed flames can be found in the supplementary material.
%Given the similar flame patterns observed during the linear phase, the data are sampled over the entire period. In contrast, during the non-linear phase, the flame structure varies significantly and becomes strongly time dependent; therefore, only samples extracted from a single snapshot are presented for illustration, taken at $t = 10\tau_{\mathrm{c}}$ for $\phi = 0.4$ and $t = 26.6\tau_{\mathrm{c}}$ for $\phi = 0.7$. 

In the non-linear phase, the growth of the instability leads to a much larger scatter in the $S_d^*$-$K$ relation. At $\phi=0.4$ conditions, two distinct branches are observed in Fig. \ref{fig7}(c). In the case of weakly stretched flame-front segments, i.e. for small values of $K$ around zero, a positive correlation between $S_d^*$ and $K$ remains evident, as expected for flames with small $Le_{eff}$. Compared with the linear phase, the increased scatter arises from the onset of strong stretch effects coupled with differential diffusion, which induces local variations in local equivalence ratio \cite{young2025influence}. Under these conditions, the assumption of a weakly stretched flame is no longer valid, and it becomes difficult to characterise the flame response using a single Markstein length. Nevertheless, a linear fit is still performed (i.e. the solid lines in Fig.~\ref{fig7}c), and the corresponding fitted slope $\mathcal{M}$ is reported for reference. For negative stretch, a negative correlation between $S_d^*$ and $K$ is observed. This behaviour appears counterintuitive, since negative stretch is generally expected to reduce the flame speed for $Le<1$ \cite{law2006combustion,chen2021effects}. However, similar behaviour has been reported in oscillating Bunsen flames \cite{yang2024effects} and is also commonly observed in turbulent premixed flames \cite{chakraborty2022assessment}. As in \cite{yang2024effects}, the negative-stretch branch is primarily associated with negative curvature generated when two branches of the wrinkled flame front approach each other during flame pinch-off. At this stage, the flame is predominantly governed by curvature-driven diffusion \cite{yang2024effects} and flame displacement speed can be described as a linear function of curvature with a positive Markstein length \cite{dave2020evolution}. This leads to $S_d^* \propto -D_c\nabla\cdot\mathbf{n} \approx -D_cK/S_d^*$, and hence $S_d^* \propto \sqrt{-D_cK}$ (see the dashed lines in Fig.~\ref{fig7}c). These two $S_d^*$-$K$ branches correspond to distinct flame topologies and are both time- and frequency-dependent. As shown earlier in Fig.~\ref{fig4}, increasing the forcing frequency promotes the development of larger-scale flame structures for $\phi=0.4$. Correspondingly, Fig.~\ref{fig7}(c) shows that the scatter becomes increaingly concentrated in the weakly-stretched region, while the number of flame segments associated with negative curvature decreases. Moreover, the stronger and more localised instabilities at higher forcing frequencies broaden the explored equivalence ratio space and enhance the flame sensitivity to stretch, thereby increasing the linear coefficient. In the case of $\phi=0.7$ flame, by contrast, an overall negative correlation between $S_d^*$ and $K$ is observed. This is due to the fact that with an $Le_{eff}$ closer to unity, the flame speed is less sensitive to stretch in the weakly-stretch regime and maintains a more uniform structure, as shown in Fig.~\ref{fig4}. As a result, the response is dominated mainly by the negative-curvature contribution. On this basis, the NQ model is applied, and the values of $\mathcal{M}$ are obtained by non-linear regression. The resulting correlation shows little sensitivity to forcing frequency, consistent with the relatively uniform flame structures observed for the $\phi = 0.7$ case across all frequencies in Fig.~\ref{fig4}.

\vspace{-4pt}
\section{Conclusions\label{conclusions}} \addvspace{10pt}
\vspace{-4pt}

In this study, 2D numerical simulations of acoustically forced lean hydrogen-air premixed flames ($\phi=0.4$ and $0.7$) are performed to examine the effects of equivalence ratio and forcing frequency on flame morphology and dynamic response. The flame evolution is found to proceed from an initially weakly stretched state to exponential perturbation growth, wrinkle interaction, and the formation of non-linear cellular structures, with distinct linear and non-linear stages identified by Fourier mode analysis. For $\phi=0.4$, the flame morphologies are strongly influenced by thermodiffusive instability, and the flame-front evolution follows a characteristic sequence of uniform cells, cell splitting, and cell merging, whereas for $\phi=0.7$ the weaker thermodiffusive effect results in a response more strongly influenced by hydrodynamic instability and large-scale wrinkle growth. At lower forcing frequencies, the flame corrugations remain relatively uniform, whereas at higher frequencies they become increasingly modulated, forming envelope-like patterns. This behaviour is interpreted as the interaction between an intrinsic standing cellular mode of the flame and the propagating acoustic perturbation. Compared to $\phi=0.7$, the $\phi=0.4$ flame exhibits stronger and more localised instabilities, because stronger differential diffusion amplifies the instability near the acoustic source and weakens acoustic transmission in the lateral direction.

In the linear regime, the $S_d^*$-$K$ relation remains approximately linear for all forcing frequencies. In the case of $\phi=0.4$, the Markstein number approaches zero as the forcing frequency increases, indicating reduced sensitivity to stretch and acoustic forcing. For the $\phi=0.7$ flame, the Markstein length remains close to zero over the entire forcing-frequency range considered. In the non-linear regime, two distinct branches are observed for $S_d^*$-$K$, corresponding to weakly stretched flame-front segments and strongly negatively curved flame segments associated with flame pinch-off. In the weakly stretched regime, a positive correlation between $S_d^*$ and $K$ remains evident for $\phi=0.4$. In the negatively stretched regime, a counterintuitive increase in flame speed with negative stretch is observed, which is dominated by negative-curvature effects as neighbouring flame fronts approach each other. As the forcing frequency increases, the stronger and more localised instabilities broaden the explored equivalence ratio space and enhance the sensitivity of the flame to stretch for $\phi=0.4$. In the case of $\phi=0.7$, the overall correlation is dominated by negatively curved flame segments and shows relatively weak sensitivity to forcing frequency.

In the future, it would be interesting to examine the effect of acoustic–flame interactions in a 3D configuration. A previous DNS study of thermodiffusively unstable laminar premixed hydrogen flames in 3D domains \cite{2024Thermodiffusively} has shown quantitative differences between 2D and 3D simulations. The quantitative differences arise primarily from the extended range of mixture fraction sampled in 3D due to stronger stretch experienced by three-dimensional flame structures. In the context of flame–acoustic interaction, it would therefore be of particular interest to investigate how the dependences of flame morphology, as well as stretch sensitivity, on frequency and equivalence ratio, are modified in 3D configurations.

\acknowledgement{CRediT authorship contribution statement} \addvspace{8pt}

{\bf XC}: Conceptualization, Writing – original draft, review \& editing, Formal analysis. {\bf FY}: Conceptualization, Writing – original draft, review \& editing, Designed the simulations. {\bf UA}: Conceptualization, Writing – review \& editing, Supervision, Funding acquisition. {\bf RSC}: Writing – review \& editing, Advised on simulations and data analysis.

\acknowledgement{Declaration of competing interest} \addvspace{8pt}

The authors declare that they have no known competing financial interests or personal relationships that could have appeared to influence the work reported in this paper.

\acknowledgement{Acknowledgments} \addvspace{8pt}

The authors are grateful for the financial support from the EPSRC (Grant: EP/Y017951/1) and the Tony Trapp Ph.D. studentship provided by Dr Tony Trapp. The computational support was provided by ARCHER2 Pioneer project (e817) and Comet HPC at Newcastle University.

\acknowledgement{Supplementary material} \addvspace{8pt}

Supplementary material is provided.

% -------------------------------------------------------------------- %
% -------------------------------------------------------------------- %
% -------------------------------------------------------------------- %
\footnotesize
\baselineskip 9pt

% -------------------------------------------------------------------- %
% -------------------------------------------------------------------- %
% -------------------------------------------------------------------- %
\clearpage
\thispagestyle{empty}
\bibliographystyle{proci}
\bibliography{PROCI_LaTeX,library}

% -------------------------------------------------------------------- %
% -------------------------------------------------------------------- %
% -------------------------------------------------------------------- %

\newpage

\small
\baselineskip 10pt

% -------------------------------------------------------------------- %
% -------------------------------------------------------------------- %
% -------------------------------------------------------------------- %

\end{document}